\documentclass{pasj00}

\begin{document}
\SetRunningHead{A.K.Inoue}{Dust-to-metal ratio in galaxies}
\Received{2003 March 31}
\Accepted{2003 June 28}

\title{Evolution of Dust-to-Metal Ratio in Galaxies}

\author{Akio K. \textsc{Inoue}\thanks{JSPS research
fellow.}\thanks{Present address: Department of Physics, Kyoto University,
Sakyo-ku, Kyoto 606-8502, akinoue@scphys.kyoto-u.ac.jp.} 
}
\affil{Department of Astronomy, Kyoto University, Sakyo-ku, Kyoto
606-8502}
\email{inoue@kusastro.kyoto-u.ac.jp}

%


\maketitle

\begin{abstract}
This paper investigates the evolution of the dust-to-metal ratio in
 galaxies based on a simple evolution model for the amount of metal and
 dust with infall. We take into account grain formation in 
 stellar mass-loss gas, grain growth by the accretion of metallic atoms
 in a cold dense cloud, and grain destruction by SNe
 shocks. Especially, we propose that the accretion efficiency is
 independent of the star-formation history. This predicts various
 evolutionary tracks in the metallicity ($Z$)--dust-to-gas ratio ($\cal D$) 
 plane depending on the star-formation history. In this framework, the
 observed linear $Z$--$\cal D$ relation of nearby spiral galaxies can be
 interpreted as a sequence of a constant galactic age. We
 emphasize that an observational study of the $Z$--$\cal D$ relation of
 galaxies at $z\sim 1$ is very useful to constrain the efficiencies of
 dust growth and destruction. We also suggest that the
 Lyman break galaxies at $z\sim 3$ have a very low dust-to-metal ratio,
 typically $\ltsim 0.1$. Although the effect of infall on the
 evolutionary tracks in the $Z$--$\cal D$ plane is quite small, the
 dispersion of the infall rate can disturb the $Z$--$\cal D$ relation
 with a constant galactic age.
\end{abstract}

\section{Introduction}

Because dust exists everywhere, stellar light is dimmed by the
interstellar dust, strong far-infrared radiation from dust is observed,
and the interstellar gas is heated/cooled by some processes involving 
dust. Moreover, comets and planets are made from dust, and various
molecules are formed on dust.
Therefore, dust is one of the most
important constituents of galaxies and the universe (e.g., \cite{whi03}). 

In the context of research on the galaxy evolution, knowledge of
the dust amount in galaxies is essential to extract the physical
properties of galaxies from observations. This is because radiation
from stars is always extinguished to some extent by internal dust
attenuation. Since it is difficult to estimate the precise amount
of attenuation from the observed photometric data \ (e.g.,
\cite{bua02}), any other information about the amount of dust is very
useful. One may think that the metallicity is a good tracer of the
amount of dust. This is true if the dust-to-metal ratio can be regarded
as universal. 

Dust grains are made from metal elements, which are injected into the
interstellar medium (ISM) by stellar mass-loss at the end of, or
during, stellar evolution. The evolution of the amount of metal in a
galaxy is traced as a consequence of the star formation activity of the
galaxy. To date, such metallicity evolution has been studied very well
since classical work in the 1970s (e.g., \cite{aud76,tin80} for reviews). 
The evolution of the amount of dust can be traced by an extension of the
evolution of metallicity if we take account of the formation,
destruction, and growth processes of dust grains in the ISM. 
After the pioneering work by \citet{dwe80}, some authors have examined
this subject (\cite{wan91,lis98,dwe98}; Hirashita~1999a,1999b,2000a;
\cite{edm01};Hirashita, et al.~2002a,2002b;\cite{hf02,mor03}).

For several nearby spiral galaxies, there is a global linear correlation
between the metallicity ($Z$) and the dust-to-gas ratio ($\cal D$)
suggested by \citet{iss90} observationally. 
This means that the dust-to-metal ratio
is nearly constant for these spiral galaxies. A simple model 
developed by \citet{hir99a} reproduces this global $Z$--$\cal D$
relation well. \citet{edm01} argues that the dust-to-metal ratio
does not change significantly during the entire galactic evolution
based on his simple model. In short, the global $Z$--$\cal D$ relation
can be interpreted as an evolutionary sequence by the model of
\citet{edm01}: 
galaxies evolve from a point of lower $Z$ and $\cal D$ to
another point of higher $Z$ and $\cal D$ with keeping a constant
dust-to-metal ratio.

On the other hand, the $Z$--$\cal D$ relation for dwarf
galaxies is not so tight; the dispersion reaches about an order of
magnitude. \citet{lis98} propose that outflows triggered by 
star-formation activity can cause such a large dispersion, whereas
Hirashita et al.~(2002b) show that the dispersion can result from a
large variety of the efficiency of dust destruction by supernovae
(SNe) shocks, even if there are no outflows; in their scenario the
variety of the efficiency is caused by an intermittent star-formation
history. In any case, the $Z$--$\cal D$ relation (or dust-to-metal
ratio) is not universal, at least for dwarf galaxies.

What determines the dust-to-metal ratio? It is a very important
question. Metal elements in the high-$z$ universe have already been 
detected (e.g., \cite{pet97}). To know the amount of dust (or
extinction), we need to know what fraction of metals is condensed into
dust.  Recently, there are several models which can be used to calculate
the spectral energy distribution of galaxies by solving the radiative
transfer in the dusty ISM  coupled with the chemical evolution model
(e.g., \cite{sil98,tak03}). In these models, a dust-to-metal ratio, such
as that in the Galaxy, is assumed. Is this really correct for dwarf
galaxies, starburst galaxies, and high-$z$ galaxies? 

In this paper, we investigate the evolution of the dust-to-metal ratio
in galaxies by using a simple model for the evolution of the amount of
dust and metal. 
Although \citet{dwe98} presents the most detailed evolution model for
the amount of metal and dust abundance for our Galaxy, we need not to
trace so the detailed abundance of metal and the composition of dust. 
The aim of this paper is to discuss on a {\it global} (or {\it mean})
dust-to-metal ratio in a galaxy. Therefore, a simple treatment, as
adopted in Hirashita's and Edmunds's work, is sufficient.

In the next section, we describe the model used in this
paper. In the model, we take account of gas (and also metal) infall in
order to discuss the time evolution on a real time-scale. Moreover, we
reconsider the efficiency of the grain-growth process in the ISM. We
then show in section 3 that the observed constancy of the dust-to-metal
ratio in nearby spiral galaxies can be interpreted as a sequence of a 
similar galactic age. The achieved conclusions in this paper are
summarized in the last section.

\section{Model}

Here, we describe the model used in this paper. First a set of
equations and meanings of notations are summarized. We next explain the
considered processes of dust formation, growth, and destruction in
the ISM. Then, a set of parameters suitable for our Galaxy is estimated.

\subsection{Equations}

First of all, the basic assumptions adopted here are summarized as
follows: (1) the ISM is treated as one zone, but two components. One is
a cold dense cloud, and the other is a warm diffuse medium. The mass
fraction of the cold cloud is fixed as a constant throughout
the whole evolution of a galaxy. (2) Stars are formed only in the cold
dense cloud following the Schmidt law of index 1. (3) The
instantaneous recycling approximation is adopted. (4) Produced metal and
dust are quickly mixed well in the ISM. (5) Dust consists of two
components. One is the refractory core, and the other is the volatile
mantle, which can exist only in a cold cloud. In other words, the
mantle dust is destroyed quickly if it goes out of the cloud. (6) No
outflow of the galactic scale.

Under the above assumptions, the time evolutions of the gas mass
($M_{\rm gas}$), the metal mass ($M_Z$), the core dust mass
($M_{\rm cor}$), and the mantle dust mass ($M_{\rm man}$) are governed
by the following equations: 
\begin{equation}
 \frac{dM_{\rm gas}}{dt}=-(1-R)\frac{\eta M_{\rm gas}}{\tau_{\rm SF}}+F\,,
\end{equation}
\begin{equation}
 \frac{dM_Z}{dt}=-(1-R)\frac{\eta M_Z}{\tau_{\rm SF}}
  +y_Z\frac{\eta M_{\rm gas}}{\tau_{\rm SF}}+Z_{\rm in}F\,,
\end{equation}
\begin{eqnarray}
 \frac{dM_{\rm cor}}{dt} &=& -\frac{\eta M_{\rm cor}}{\tau_{\rm SF}}
  -\frac{M_{\rm cor}}{\tau_{\rm SN}} 
  +f_{\rm c}\left(R\frac{\eta M_Z}{\tau_{\rm SF}} 
	     +y_Z\frac{\eta M_{\rm gas}}{\tau_{\rm SF}}\right) \nonumber \\
  && +\xi\frac{M_{\rm cor}(1-\delta_{\rm cl})}{\tau_{\rm acc}}\,,
\end{eqnarray}
\begin{equation}
 \frac{dM_{\rm man}}{dt}=-\frac{M_{\rm man}}{\tau_{\rm SF}}
  +(1-\xi)\frac{M_{\rm cor}(1-\delta_{\rm cl})}{\tau_{\rm acc}}\,,
\end{equation}
where $\tau_{\rm SF}$, $\tau_{\rm SN}$, and $\tau_{\rm acc}$ are the
time scales of the star formation, dust destruction by SNe, and dust
growth by metal accretion in the cold cloud, respectively. 
The term $\eta M_{\rm gas}/\tau_{\rm SF}$ is the star-formation rate,
where $\eta$ is the mass fraction of the cold gas.
The notations are summarized in table 1 and are described in the rest of
this subsection. In section 3, we will solve equations (1)---(4) combined
with the initial condition, $M_{\rm gas}=M_Z=M_{\rm cor}=M_{\rm
man}=0$ at $t=0$.

\begin{table}
  \caption{Notations and standard set.}
  \begin{center}
    \begin{tabular}{ccc}\hline\hline
     Parameter & Value & Meaning\\ \hline
     $\tau_{\rm SF}$ & 5 Gyr & star formation time-scale\\
     $\beta_{\rm in}$ & 0.5 & infall parameter ($\equiv \tau_{\rm SF}/\tau_{\rm in}$)\\
     $R$ & 0.299 & returned mass fraction\\
     $y_Z$ & 0.025 & stellar yield\\
     $Z_{\rm in}$ & 0.01$Z_\odot$ & metallicity in infall gas\\
     $f_{\rm c}$ & 0.1 & condensation efficiency\\
     $\beta_{\rm SN}$ & 10 & SNe parameter ($\equiv \tau_{\rm SF}/\tau_{\rm SN}$)\\
     $\tau_{\rm acc,0}$ & $10^8$ yr & accretion growth time-scale\\
     $\eta$ & 0.5 & cold gas fraction\\
     $\xi$ & 0.5 & refractory probability\\
     \hline
    \end{tabular}
  \end{center}
\end{table}

We consider gas and metal infall from a halo, which can resolve the
so-called G-dwarf problem of the stellar metallicity distribution in the
Galaxy (e.g., \cite{twa80}). Here, the gas infall rate is denoted by
$F$, which is assumed to be the following analytic function: 
\begin{equation}
 F=\frac{M_0}{\tau_{\rm in}} \exp[-t/\tau_{\rm in}]\,,
\end{equation}
where $\tau_{\rm in}$ is the time-scale of infall and $M_0$ is the total
gas mass that can fall into the disk from the halo i.e., $\int F
dt=M_0$. The metallicity in the infalling gas is denoted as $Z_{\rm in}$.

For the chemical evolution part [equations (1) and (2)], we need the
returned mass fraction, $R$, and the stellar yield, $y_Z$.
The returned fraction is the mass fraction of returned gas from stars
into the ISM per unit mass of star formation, defined as 
\begin{equation}
 R=\int_{m_{\rm t}}^{m_{\rm up}} (m-w_m) \phi (m) dm\,,
\end{equation}
where $w_m$ is the remnant mass of a star with mass $m$, $m_{\rm t}$ is
the present turn-off mass, and $\phi (m)$ is the initial mass function
(IMF), normalized as 
\begin{equation}
 \int_{m_{\rm low}}^{m_{\rm up}} m \phi (m) dm=1 \,,
\end{equation}
where $m_{\rm low}$ and $m_{\rm up}$ are the lower and upper cutoff
masses of the IMF. When we assume the Salpeter IMF with $m_{\rm
low}=0.1\MO$, $m_{\rm up}=100\MO$, and $m_{\rm t}=1\MO$, and assume the
remnant mass function in \citet{pag97}, we obtain $R=0.299$.
The stellar yield is the mass fraction of the metal elements newly
produced and ejected into the ISM per unit star formation, and defined as 
\begin{equation}
 y_Z=\int_{m_{\rm t}}^{m_{\rm up}} mp_Z(m)\phi(m)dm\,,
\end{equation}
where $p_Z(m)$ is the fractional mass of metal elements ejected
into the ISM by a star of mass $m$. Assuming the Salpeter IMF with the
same characteristic masses above and $p_Z(m)$ in \citet{pag97}, we
obtain $y_Z=0.025$.

In the part concerning the evolution of the amount of dust [equations
(3) and (4)], $f_{\rm c}$
is the condensation efficiency in the stellar mass-loss gas, and
$\delta_{\rm cl}$ is the dust-to-metal ratio in a cold cloud, i.e., 
\begin{equation}
 \delta_{\rm cl}=\frac{M_{\rm cor}}{M_Z}
  +\frac{M_{\rm man}}{\eta M_Z}\,.
\end{equation}
Finally,  $\xi$ is the refractory probability of metal
accretion. The refractory core may grow by an 
accretion process in the cold cloud as well as the volatile mantle.
This parameter is defined as the fraction of accreted metal mass
which contributes to the growth of the refractory core.

\subsection{Grain Formation, Evolution, and Destruction}

We first consider the condensation efficiency in the stellar outflow,
$f_{\rm c}$. 
The efficiency may depend on the stellar mass; different
efficiencies between that in mass ejection by SNe and that in stellar wind
from evolved stars are possible. In the framework presented here, we
do not distinguish these two types of stellar mass-loss because of
the instantaneous recycling approximation. Thus, the condensation
efficiency given in this paper is an effective one, including both types
of mass loss and averaged over the stellar mass by the assumed IMF. 
Moreover, the efficiency can evolve with the stellar
metallicity. Indeed, \citet{mor03} suggest such an evolution, but
the evolution is not very significant, fortunately. This is because the
metal amounts in stellar outflows are often dominated by the newly
produced metal. Therefore, we neglect the metallicity dependence of
$f_{\rm c}$, and assume a constant $f_{\rm c}$.

The ``stardust'' injection rate is $f_{\rm c}(RZ+y_Z)\eta
M_{\rm gas}/\tau_{\rm SF}$ [i.e., third term in equation (3)], where $Z$
is the metallicity ($=M_Z/M_{\rm gas}$). 
On the other hand, the stellar mass-loss rate is 
$R\eta M_{\rm gas}/\tau_{\rm SF}$. 
Again, we note that the injection
rate and the mass-loss rate include all types of stellar outflows. 
Thus, the dust-to-gas ratio in the mass-loss gas (${\cal D}_*$) is 
\begin{equation}
 {\cal D}_*=\frac{f_{\rm c}(RZ+y_Z)}{R}
  \approx f_{\rm c}\frac{y_Z}{R}\,.
\end{equation} 
The approximation in the last term is valid when $Z \ll 4Z_\odot$ for
$R=0.299$, $y_Z=0.025$, and the solar metallicity $Z_\odot =0.02$.
Observationally, ${\cal D}_* \sim 0.01$ is suggested as the present-day
Galactic value \citep{geh89}, although the uncertainty is still
large. Therefore, we obtain $f_{\rm c} \sim 0.1$, which agrees
with values given by Hirashita~(1999a,1999b) and \citet{edm01}.

Next, we discuss the grain growth rate in the cold cloud. The dust mass
increases if the metallic atoms accrete on the pre-existing grain
cores. 
The refractory core, itself, may grow by accretion as well as
the volatile mantle. If the accreted metal elements are refractory, 
such as Si, Fe, or Ti, the refractory core grows. On the other hand,
elements such as C, N, and O produce a volatile mantle (maybe ice
molecules) on the refractory core when these elements accrete onto the
grain surface. Actually, the accretion rates of various elements are
different from each other. To avoid such difficulty, we parameterize the
situation by the refractory probability, $\xi$. This is defined as the
mass fraction of accreted metal contributing to the growth of the
refractory core. 

The total (core+mantle) dust mass growth rate by metal accretion can
be expressed as   
\begin{equation}
 \left[\frac{dM_{\rm d}}{dt}\right]_{\rm acc} 
  = N_{\rm cor}^{\rm cl} \pi a^2 s_Z \rho_Z^{\rm cl} 
  \langle v_Z \rangle\,,
\end{equation}
where $N_{\rm cor}^{\rm cl}$ is the number of refractory cores in the
cold clouds, $a$ is the typical grain radius, $s_Z$ is the mean 
sticking coefficient of metal elements, $\rho_Z^{\rm cl}$ is the
mass volume density of gaseous metallic atoms in the cold clouds, and
$\langle v_Z \rangle$ is the mean velocity of metallic atoms.
We assumed a spherical grain for simplicity. 
Since $N_{\rm cor}^{\rm cl}=\eta M_{\rm cor}/m_{\rm d}$
and $\rho_Z^{\rm cl}=\rho_{\rm gas}^{\rm cl}Z(1-\delta_{\rm cl})$,
where $m_d$ is the typical grain mass and $\rho_{\rm gas}^{\rm cl}$ is
the volume mass density of the cold cloud gas, the dust mass growth rate
is reduced to  
\begin{equation}
 \left[\frac{dM_{\rm d}}{dt}\right]_{\rm acc} 
  = \frac{M_{\rm cor}(1-\delta_{\rm cl})}{\tau_{\rm acc}}\,,
\end{equation}
where we define the accretion time scale as $\tau_{\rm acc}=m_d/\eta \pi
a^2 s_Z \rho_{\rm gas}^{\rm cl} \langle v_Z \rangle Z$.
We then obtain the fourth term in equation (3) and the second term in
equation (4) by using the refractory probability, $\xi$. The parameter
is determined later so as to reproduce the Galactic metallicity and
dust-to-gas ratio.  

The accretion time-scale, $\tau_{\rm acc}$, is not constant
during galaxy evolution, but depends on the metallicity. Here, 
we normalize it as 
\begin{equation}
 \tau_{\rm acc}=\tau_{\rm acc,0}(Z_\odot/Z)\,, 
\end{equation}
and we estimate the normalization, $\tau_{\rm acc,0}$, which means the
accretion time-scale at the solar metallicity (i.e., the current value
in the Galaxy). According to \citet{hir00b}, who estimated the grain
growth time-scale by taking into account the mass function of the
molecular clouds in the Galaxy, 
$\tau_{\rm acc,0} \sim \eta^{-1} 5 \times 10^7$ yr.  
Although the cold gas fraction, $\eta$, is rather uncertain, $\eta$
may not deviate significantly from 0.5. Thus, we always assume
$\tau_{\rm acc,0} \sim 10^8$ yr, except when indicated otherwise. 

We next consider the dust-destruction rate. There are two processes for
destruction; one is incorporation by star formation; the
other is evaporation and sputtering in SNe remnants. Here, we assume
that star formation occurs only in a cold cloud. Since
the dust-to-gas ratio in a cold cloud is 
$(\eta M_{\rm cor} + M_{\rm man})/\eta M_{\rm gas}$, 
and the star-formation rate is $\eta M_{\rm gas}/\tau_{\rm SF}$, we
obtain the first terms in equations (3) and (4) as the destruction rate
by star formation of core and mantle, respectively. Here, we have assumed
the destruction efficiency by star formation to be unity.

The SNe destroy only the core dust because the mantle dust is already
destroyed in the SNe remnants, which are not cold gas. In fact, the
SNe affect the amount of mantle dust through the cold gas
fraction, which is reduced by the SNe. However, we assume a constant
fraction of cold gas for simplicity. Since the dust-to-gas ratio in
a warm medium is $M_{\rm cor}/M_{\rm gas}$, the destruction rate by
SNe is 
\begin{equation}
 \left[\frac{dM_{\rm d}}{dt}\right]_{\rm SN} 
  = - \frac{M_{\rm cor}}{M_{\rm gas}} \epsilon M_{\rm SNR} \gamma\,,
\end{equation}
where $\epsilon$ is the destruction efficiency, $M_{\rm SNR}$ is the gas
mass shocked by one SN, and $\gamma$ is the occurrence rate of SNe.
We do not distinguish type Ia and type II SNe.
Thus, the SNe-rate $\gamma$ means the sum of the occurrence rates 
of both type Ia and type II.
If we define the time-scale of the destruction by SNe as $\tau_{\rm
SN}=M_{\rm gas}/\epsilon M_{\rm SNR} \gamma$ \citep{mck89}, the second
term in equation (3) is obtained. 

\citet{dwe98} proposed $\tau_{\rm SN}/\tau_{\rm acc,0}\simeq 2$
(constant). It was also assumed in Hirashita (1999a,b). Here, we
reconsider this argument.  When the star-formation rate is denoted as
$\psi$($=\eta M_{\rm gas}/\tau_{\rm SF}$), we find 
$\tau_{\rm SN}/\tau_{\rm SF}=(\psi/\gamma)/(\eta \epsilon M_{\rm SNR})$. 
On the other hand, we have seen that $\tau_{\rm acc,0} \propto \eta^{-1}$. 
Thus, we find $\tau_{\rm SN}/\tau_{\rm acc,0} \propto (\psi/\gamma)
\tau_{\rm SF}$. We may consider that the ratio of $\psi$ and $\gamma$ is
roughly constant. Therefore, the ratio $\tau_{\rm SN}/\tau_{\rm acc,0}$
is proportional to $\tau_{\rm SF}$. In other words, the ratio is not
constant, as assumed by \citet{dwe98}, but dependent on the
star-formation history (i.e., $\tau_{\rm SF}$). This provides us with a
new interpretation of the $Z$--$\cal D$ relation for nearby spiral
galaxies presented in the next section. It is worth noting that we do
not need the assumption of a constant $\eta$ in the above statement.

Although the value of $\eta$ is required when we
estimate $\tau_{\rm acc,0}$ and $\tau_{\rm SF}$, estimating $\eta$ is a
rather complex task.  
After stars are formed and SNe occurs, the
cold gas is heated by the SNe shocks, so that $\eta$ decreases. When
$\eta$ decreases, the star-formation rate is suppressed, and then the SNe
occurrence rate decreases, so that $\eta$ increases. Thus, the
star-formation rate ($\propto \tau_{\rm SF}^{-1}$) and $\eta$ are coupled
with each other non-linearly \citep{ike83,hir00a}. To avoid the complex
task of solving a set of non-linear equations to determine $\eta$, we
assume a constant $\eta$ as an average value during the galaxy evolution
time-scale for simplicity. For a constant $\eta$, the ratio of
$\tau_{\rm SN}$ to $\tau_{\rm SF}$ becomes a constant. Since
\citet{mck89} estimates $\tau_{\rm SN} \simeq 4\times 10^8$ yr for the
Galaxy and $\tau_{\rm SF} \sim 5\times 10^9$ yr is suitable for the
Galaxy, we obtain $\tau_{\rm SF}/\tau_{\rm SN} \sim 10$. Combining
$\tau_{\rm acc,0} \sim 10^8$ yr, we find 
\begin{equation}
 \frac{\tau_{\rm SN}}{\tau_{\rm acc,0}}
  \sim \left(\frac{\tau_{\rm SF}}{1~{\rm Gyr}}\right)\,.
\end{equation}
In future work, the fraction $\eta$ should be determined
self-consistently in the model.

\subsection{Standard Case (Milky Way)}

\begin{figure}
  \begin{center}
    \FigureFile(80mm,80mm){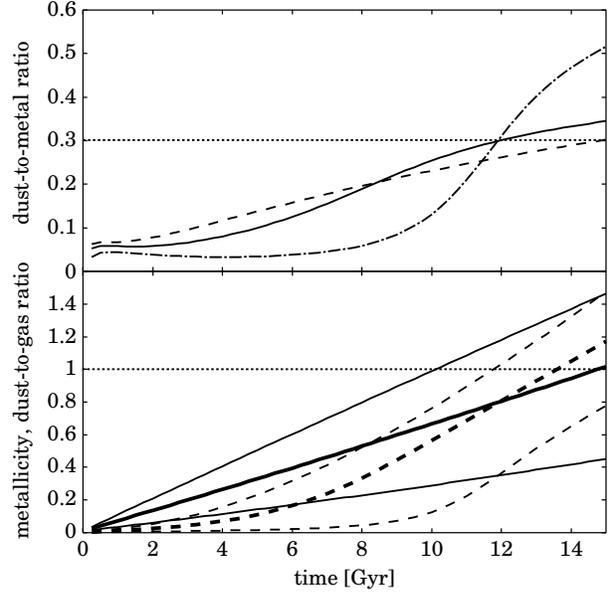}
  \end{center}
  \caption{Time evolution of the metallicity, dust-to-gas ratio, 
 and dust-to-metal ratio. {\it Bottom} : the solid curves
 indicate the metallicity and the dashed curves indicate the dust-to-gas
 ratio. The vertical axis is normalized by the present-day Galactic
 value, which is indicated by the dotted straight line. The thick curves
 are the standard case (table 1), the top thin (solid and dashed) curves are
 the case of $\eta=0.8$ and $\xi=0$, and the bottom thin curves are the case of
 $\eta=0.2$ and $\xi=1$, where $\eta$ is the cold gas fraction and $\xi$
 is the refractory probability in the accretion process. {\it Top} : the
 solid, dashed, and dash-dotted curves are the standard case, $\eta=0.8$
 and $\xi=0$ case, and $\eta=0.2$ and $\xi=1$ case, respectively. The
 dotted straight line indicates the present-day Galactic value.}
\end{figure}

Here, we estimate a set of parameters suitable for our Galaxy. 
First, the time-scales of the star formation and infall are discussed.  
\citet{tak00} have examined the star-formation history of the Galaxy by
using a chemical evolution model with infall similar to that used in
this paper. According to them, a set of these time-scales between 
$(\tau_{\rm SF},\tau_{\rm in})=(6,23)$ and (11,12) in Gyr unit can
reproduce the observed star-formation history of the galactic disk,
the age--metallicity relation of stars in the solar neighborhood, and
the metallicity distribution of G-dwarfs. Noting that their star-formation
rate and that in this paper are different in the treatment of the
cold-gas fraction ($\eta$), we adopt $\tau_{\rm SF}=5$ Gyr 
and $\tau_{\rm in}=10$
Gyr. Combining these time-scales with $R=0.299$ and $y_Z=0.025$
(see subsection 2.1), and assuming a very low metallicity in the infall gas
(for example $Z_{\rm in}=0.01Z_\odot$), we find that the present-day
(age $\sim 12$ Gyr) metallicity becomes around Solar metallicity (see
figure 1 bottom panel), which ensures the validity of these time-scales. 

For the convenience of later discussions, we define the infall
parameter, $\beta_{\rm in}$, as $\tau_{\rm SF}/\tau_{\rm in}$. Thus,
we adopt $\beta_{\rm in}=0.5$ for our Galaxy. Since an infall parameter
suitable for other galaxies with a different star-formation history
(i.e., $\tau_{\rm SF}$) is rather unknown, we assume $\beta_{\rm
in}=0.5$ for all galaxies. The effect of the other choices of $\beta_{\rm
in}$ is discussed in subsection 3.2.

We use values estimated in subsection 2.2 for parameters $f_{\rm c}$,
$\beta_{\rm SN}\equiv \tau_{\rm SF}/\tau_{\rm SN}$, and $\tau_{\rm
acc,0}$. Strictly speaking, $\tau_{\rm acc,0}$ depends on $\eta$, as
shown in subsection 2.2. 
Nevertheless, we adopt a constant $\tau_{\rm acc,0}=10^8$
yr for simplicity because the uncertainty of $\tau_{\rm acc,0}$ is large
and $\eta$ may be roughly $\sim 0.5$.

We now modulate two parameters of $\eta$ (cold gas fraction) and $\xi$
(refractory probability in accretion) so as to reproduce the present-day
dust-to-metal ratio in the Galaxy. Adopting the solar metallicity (0.02)
and the dust-to-gas ratio in the solar neighborhood ($6\times 10^{-3}$; 
\cite{spi78}) as the present-day $Z$ and $\cal D$ in the Galaxy, we find
the present-day dust-to-metal ratio, $\delta=(M_{\rm cor}+M_{\rm
man})/M_{\rm gas}=0.3$. In figure 1, we show the time evolution of the
metallicity, the dust-to-gas ratio, and the dust-to-metal ratio for
three sets of $\eta$ and $\xi$: 
(1) $\eta=\xi=0.5$, (2) $\eta=0.8$ and $\xi=0$, and
(3) $\eta=0.2$ and $\xi=1$. All of these cases give the calculated
$\delta\simeq 0.3$ at the galactic age of 12--15 Gyr (top panel in
figure 1). From the bottom panel of figure 1, we find that the case of
$\eta=\xi=0.5$ is the best. Thus, we call this case the 'standard case', 
whose parameters are summarized in table 1. 

Although the case of $\eta=0.2$ and $\xi=1$ is not suitable for the
Galaxy, because of the too small metallicity, 
the case of $\eta=0.8$ and $\xi=0$ is not so bad. Since the time
evolution of the dust-to-metal ratio in this case is very similar to
that of the standard case, our results obtained later are not altered if
we choose this case as the standard case. We note here that the
parameter set given in table 1, as the standard set, is not a unique
solution for the Galaxy. There may be other sets which can reproduce all
observational constraints for the Galaxy. However, we do not have enough
data for determining a unique solution, even for the Galaxy. For example,
we can determine $\xi$ if we know the mass ratio of the grain core and
mantle, which is still unknown. In future work, we may determine
these parameters more precisely.

It is important that the dust-to-metal ratio in young galaxies can be
much smaller than that in the present-day Galaxy, as shown in the top
panel of figure 1. Therefore, a galaxy does not evolve with a constant
dust-to-metal ratio. That is, we cannot consider the $Z$--$\cal D$
relation of Issa et al.~(1990) to be an evolutionary sequence, like
\citet{edm01}. We discuss what the $Z$--$\cal D$
relation implies in the next section.

\section{Results and Discussions}

\subsection{$Z$--$\cal D$ Relation}

Here, we compare the observed $Z$--$\cal D$ relation with our theoretical
ones. First, the observational data used are explained. Although a large
and systematic data set of $Z$ and $\cal D$ of various types of galaxies
is required, we cannot find such a data set. Thus, we use
the data presented by Issa et al.~(1990), whose sample galaxies are five
nearby spiral galaxies (the Galaxy, M 31, M 33, M 51, and M 101) and both
Magellanic clouds. \citet{lis98} and Hirashita et al.~(2002b) compiled such
data of dwarf galaxies. However, since their $\cal D$s of dwarf galaxies
are based only on IRAS data, there may be systematic
underestimations due to the cold dust not detected by IRAS (e.g.,
\cite{pop02}). Thus, we decided to use only Issa's data. For M 31, the
dust-to-gas ratio estimated in \citet{ino01} from the data obtained
by ISO was also used. It is very useful to examine the $Z$--$\cal
D$ relation for a large sample of galaxies in future work.

\begin{figure}
  \begin{center}
    \FigureFile(80mm,80mm){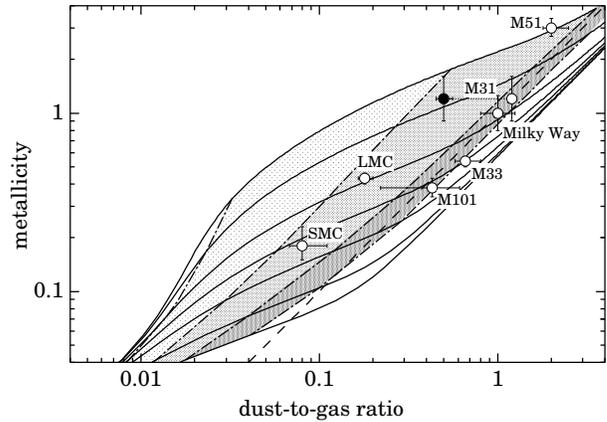}
  \end{center}
  \caption{Metallicity ($Z$)--dust-to-gas ratio ($\cal D$) relation. The
 vertical and horizontal axes are normalized by $Z_\odot$ and $\cal D$
 of the Galaxy, respectively. The open circles are observed data of $Z$
 and $\cal D$ for five nearby spiral galaxies and Magellanic clouds taken
 from Issa et al.~(1990), except for M 31 whose $\cal D$ is taken from
 \citet{ino01}. The filled circle is M 31 taken from Issa et
 al.~(1990). The solid curves are theoretical $Z$--$\cal D$ relations;
 the adopted star formation time-scale $\tau_{\rm SF}=1$, 2, 5, 10, 20,
 50, and 100 Gyr from top to bottom. Other parameters are the same as
 those in table 1. The dash-dotted curves indicate the iso-galactic age; 1,
 5, 10, and 15 Gyr from left to right. That is, the thin, medium, and
 thick shaded areas indicate that the galactic age is in 1--5 Gyr, 5--10
 Gyr, and 10--15 Gyr, respectively. The dashed straight line represents
 the sequence of the same dust-to-metal ratio as the Galaxy.}
\end{figure}

In figure 2, we display a comparison between the observed data points
and the theoretical predictions with various star-formation histories 
(i.e., $\tau_{\rm SF}$). The adopted $\tau_{\rm SF}$s are 1, 2, 5, 10, 20,
50, and 100 Gyr from top to bottom. They are suitable for early- to
late-type galaxies from top to bottom. In our model, a different
evolutionary path in the $Z$--$\cal D$ plane is predicted depending
on $\tau_{\rm SF}$. This is because $\tau_{\rm SN}/\tau_{\rm acc,0}$ is
not constant, but dependent on $\tau_{\rm SF}$, as discussed in subsection
2.2 [equation (15)]. In other words, the accretion time-scale does not
depend on $\tau_{\rm SF}$, whereas the SNe destruction time-scale
depends on it. The separation of these evolutionary sequences is
caused by the different efficiency of the dust destruction by SNe
relative to the accretion growth of dust. A more rapid star formation
causes a higher occurrence rate of SNe, and results in a more efficient
destruction of dust.

How can we understand the observed constancy of the dust-to-metal
ratio in nearby spiral galaxies within the framework of our model? 
The answer is that the sequence of the dust-to-metal ratio is the
sequence of a constant galactic age. As shown in figure 2, these spiral
galaxies are plotted in (or around) the area with an age of 10--15 Gyr
(thick shaded area). Therefore, we conclude that the observed linear
$Z$--$\cal D$ relation for the nearby spiral galaxies is not an 
evolutionary sequence with a constant dust-to-metal ratio, but 
a sequence of a similar galactic age with various star-formation
histories (i.e., from early- to late-type). 
This is the main conclusion of this paper. 

\begin{figure}
  \begin{center}
    \FigureFile(80mm,80mm){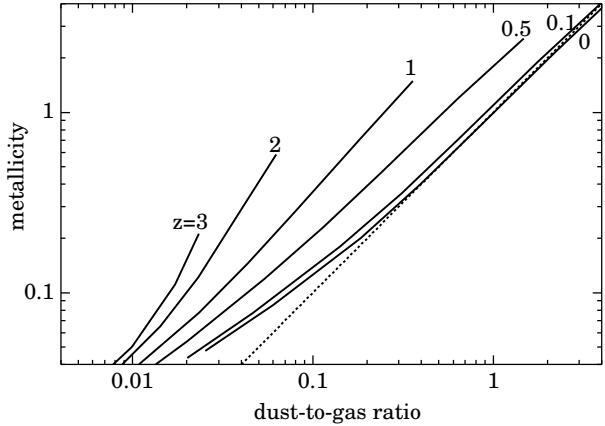}
  \end{center}
  \caption{Predicted redshift evolution of $Z$--$\cal D$ relation. The
 solid curves are the predicted $Z$--$\cal D$ relation at $z=3$, 2, 1,
 0.5, 0.1, and 0 (local) from left to right, as indicated in the
 panel. The dotted straight line represents the sequence of the same
 dust-to-metal ratio as the Galaxy. The vertical and horizontal axes are
 the same mean as those of figure 2. The present-day galactic age is
 assumed to be 12 Gyr. The adopted cosmology is $\Omega_{\rm M}=0.3$,
 $\Omega_\Lambda=0.7$, and $H_0=70$ km s$^{-1}$ Mpc$^{-1}$. The
 vertical and horizontal axes are normalized by $Z_\odot$ and $\cal D$
 of the Galaxy, respectively.} 
\end{figure}

As shown in figure 1, our model predicts a time evolution of the
dust-to-metal ratio: a lower dust-to-metal ratio in younger
galaxies. Thus, the $Z$--$\cal D$ relation is expected to evolve with
the galactic age or redshift. We show such redshift evolution of the
$Z$--$\cal D$ relation in figure 3, where we assume the galactic age at
the present-day to be 12 Gyr. We predict a deviation from the local
$Z$--$\cal D$ relation at $z>0.5$. Therefore, the investigation of the
$Z$--$\cal D$ relation in a high-$z$ will be a good test against our
model. Observations with, for example, ASTRO-F and SIRTF will be
very useful.

Finally, we comment on the Magellanic clouds. They are plotted in
a somewhat younger area in figure 2 than other spiral galaxies. They are
found in the area of an age of 5--10 Gyr (medium shaded area). 
The definition of the galactic age in the model is the time from the
start of the main gas infall and star formation because our
treatment is one-zone. If a minor star formation exists before the main
one, we underestimate the galactic age. Thus, the younger ages of
Magellanic clouds may be due to such an effect.

\subsection{Effect of SNe, Accretion, and Infall Time-Scales}

In this subsection we discuss the effect on our interpretation of the
local $Z$--$\cal D$ relation when we change $\tau_{\rm SN}$, 
$\tau_{\rm acc,0}$, and $\tau_{\rm in}$. 
\citet{edm01} argues that the accretion time scale is much shorter than
that adopted here, and that the dust destruction by SNe is not efficient. 
Indeed, he assumes the sticking probability on the
grain surface in the accretion process of the metallic atoms to be unity,
whereas our estimate is based on the value of \citet{hir00b}, who assumes 
a sticking probability of 0.1. The accretion time scale becomes
an order of magnitude shorter than that in table 1 if we assume a 
sticking probability of unity. 

\begin{figure}
  \begin{center}
    \FigureFile(80mm,80mm){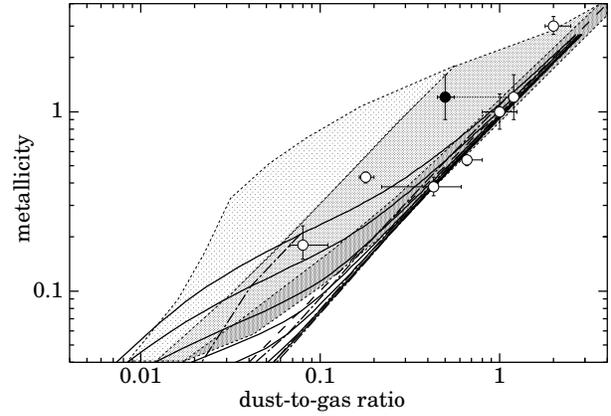}
  \end{center}
  \caption{Same as figure 2, but $\tau_{\rm acc,0}=10^7$ yr and
 $\eta=0.3$ are adopted. The thin, medium, and thick shaded areas
 indicate the same areas as in figure 2 for a comparison. The
 vertical and horizontal axes are normalized by $Z_\odot$ and $\cal D$
 of the Galaxy, respectively.}
\end{figure}

In figure 4, the high accretion cases are displayed. All curves and
points, except for the shaded areas, have the same meanings as in figure
2, but $\tau_{\rm acc,0}=10^7$ yr and $\eta=0.3$ are adopted.
The value of $\eta$ is determined so as to reproduce the
parameters of the Galaxy, as done in subsection 2.3. 
The shaded areas cover the same areas as figure 2 for a 
comparison. We find that the variation of the $Z$--$\cal D$ relations 
becomes small and the evolutionary tracks of various $\tau_{\rm SF}$ are
almost superposed on the dashed straight line, which is a sequence of
the same dust-to-metal ratio as the Galaxy if the galactic age is larger
than about 5 Gyr. This means that the local $Z$--$\cal D$ relation of
spiral galaxies can be interpreted as the evolutionary sequence. Of
course, we can interpret it as a sequence of similar galactic age, 
although sequences of age larger than about 5 Gyr cannot be
distinguished from each other. 

\begin{figure}
  \begin{center}
    \FigureFile(80mm,80mm){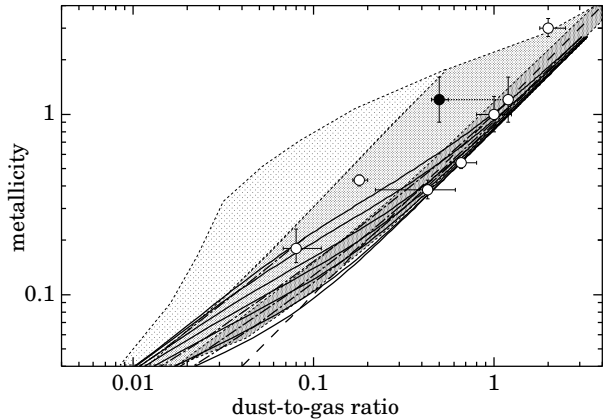}
  \end{center}
  \caption{Same as figure 2, but $\beta_{\rm SN}= \tau_{\rm SF}/\tau_{\rm SN} 
 =0$, $\eta=0.3$, and $\xi=0$ are adopted. The thin, medium, and thick
 shaded areas indicate the same areas as in figure 2 for a comparison. The
 vertical and horizontal axes are normalized by $Z_\odot$ and $\cal D$
 of the Galaxy, respectively.}
\end{figure}

In figure 5, we display the case of no SNe destruction. All curves and
points, except for the shaded areas, have the same meanings as in figure
2, but $\beta_{\rm SN}=0$, $\eta=0.3$, and $\xi=0$ are adopted.
The values of $\eta$ and $\xi$ are adjusted so as to reproduce the
Galactic parameters. The shaded areas indicate the same areas as in figure
2 for a comparison. We find a similar trend to the high accretion
case, a tighter $Z$--$\cal D$ relation. We can again interpret the local
$Z$--$\cal D$ relation as the evolutionary sequence. 

Although the data point of the Large Magellanic Cloud cannot be covered
by the no SNe destruction and high accretion unless $\tau_{\rm SF}<1$
Gyr, this may not be so critical because the uncertainty of the data
point is likely to be larger than its error-bars shown in figures 2, 4,
and 5.
Therefore, we cannot distinguish the standard, no SNe destruction, and
the high-accretion cases only by the local $Z$--$\cal D$ relation. However,
fortunately, these three cases can be distinguished if we have 
information about the $Z$--$\cal D$ relation of younger galaxies.

\begin{figure}
  \begin{center}
    \FigureFile(80mm,80mm){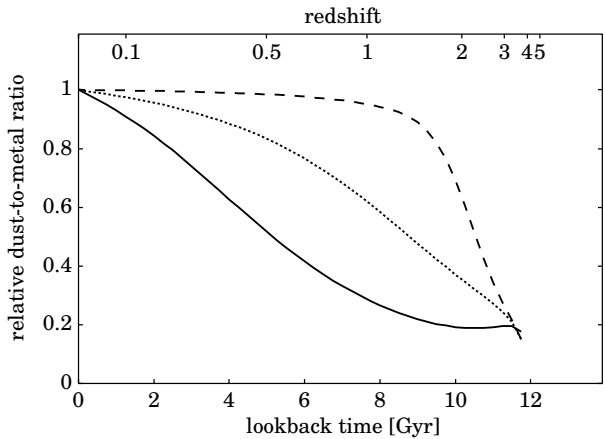}
  \end{center}
  \caption{Time evolution of the dust-to-metal ratio. The vertical axis is
 normalized by the present-day value. The solid, dotted, and dashed
 curves are the standard (table 1), no SNe destruction, and high
 accretion cases, respectively. For all curves, $\tau_{\rm SF}=5$ Gyr and
 the present-day galactic age of 12 Gyr are assumed. The adopted
 cosmology is $\Omega_{\rm M}=0.3$, $\Omega_\Lambda=0.7$, and $H_0=70$
 km s$^{-1}$ Mpc$^{-1}$.}
\end{figure}

In figure 6, we show the dust-to-metal ratios of the standard, no SNe
destruction, and high accretion as a function of the lookback time and
redshift. A present-day galactic age of 12 Gyr and $\tau_{\rm SF}=5$
Gyr are assumed. From this figure, we find that the dust-to-metal ratios
of the standard and no SNe destruction cases at $z=1$ become 30\% and
60\% of the local value, respectively, whereas that of the
high-accretion case does not change significantly. Thus, we can obtain
further constraints on $\tau_{\rm SN}$ and $\tau_{\rm acc,0}$ by using
observations of the $Z$--$\cal D$ relation at $z\sim 1$ 
(or local galaxies with an age of several Gyr).

\begin{figure}
  \begin{center}
    \FigureFile(80mm,80mm){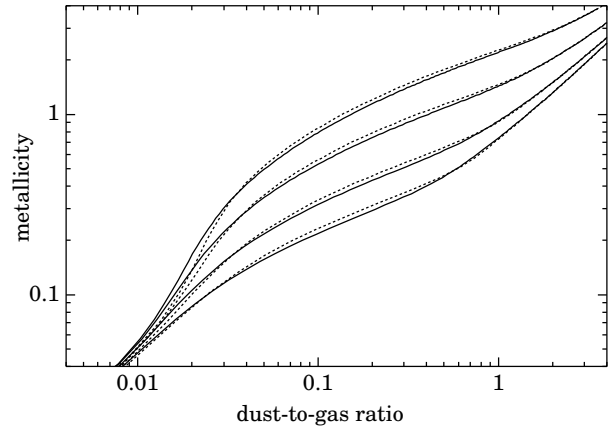}
  \end{center}
  \caption{Effect of infall on the $Z$--$\cal D$ relation. The solid and
 dotted curves are the $\beta_{\rm in}=0.5$ and no infall cases,
 respectively. The adopted $\tau_{\rm SF}=1$, 2, 5, and 10 Gyr from top
 to bottom. The vertical and horizontal axes are normalized by $Z_\odot$
 and $\cal D$ of the Galaxy, respectively.}
\end{figure}

The effect of infall on the $Z$--$\cal D$ relation is
discussed. In Figure 7, we show the comparison between the standard
cases and the no infall cases. Clearly we find the effect of infall is
negligible in the plane of $Z$ and $\cal D$. However, the infall
parameter $\beta_{\rm in}$ affects the speed of the chemical
enrichment. Thus, a galaxy without infall rapidly evolves in the
$Z$--$\cal D$ plane along the track determined from the combination of
parameters, $\tau_{\rm acc,0}$, $\beta_{\rm SN}$, $\eta$, and $\xi$.
Therefore, the dispersion of $\beta_{\rm in}$ can disturb the
$Z$--$\cal D$ relation of a constant galactic age as shown in Figure 3.

\subsection{Implication for Very Young Galaxies}

All our models considered in figure 6 predict that the
dust-to-metal ratio in a galaxy younger than about 1 Gyr is very small,
about one-fifth of the present-day Galactic value (dust/metal $\ltsim
0.1$). Because of the instantaneous recycling approximation, the
accuracy of the model result may not be so good for such very young
galaxies. Nevertheless, we discuss what our model implies against 
very young galaxies whose age is less than 1 Gyr.

First, we compare our model prediction with observations of a 
nearby dwarf galaxy, SBS 0335$-$052. This dwarf galaxy has an age of less
than about 100 Myr (\cite{izo97}). In figures 8 and 9, we show such
comparisons. The allowed range of the metallicity is taken from
\citet{izo97}. That of the dust-to-gas ratio is based on the infrared
spectral energy distribution (\cite{dal01}) and the hydrogen Brackett
lines ratio (\cite{hun01}). For theoretical
predictions, we assume $\tau_{\rm SF}=1$ (thick curves) and 5
(thin curves) Gyr. As can be seen in figure 8, the effect of the
metallicity in the infalling gas, which is assumed to be $0.01 Z_\odot$,
appears in the solid ($\eta=0.5$) and dashed ($\eta=0.3$) curves. The
dash-dotted and dotted curves are the cases of zero metal infall and no
infall, respectively. From these figures, we find that our model
reproduces the observed metallicity, dust-to-gas ratio, and very young
age quite well if we assume $\tau_{\rm SF}=1$ Gyr. Such a short
time-scale of the star formation, i.e., a starburst like time-scale, is
reasonable for SBS 0335$-$052 because the galaxy is really starbursting
now. 

Next, we move on a comparison with the Lyman break galaxies (LBGs) at
$z \sim 3$. The LBGs seem to have an age from about 100 Myr to 1 Gyr
(e.g., \cite{sha01}) and a metallicity around one-third of the Solar
value (e.g., \cite{pet01}). The dust-to-gas ratio in the LBGs is still
quite uncertain.  We compare only the metallicity of the LBGs with our
model. figure 8 shows that our models with $\tau_{\rm SF}=1$ Gyr
reproduce the observed metallicity and age of the LBGs very well. In
figure 9, we present the expected range of the dust-to-gas ratio in the
LBGs (the thin shaded area); 0.01--0.1 of the dust-to-gas ratio of our
Galaxy. Our model predicts that the dust-to-metal ratio in the LBGs is
$\sim 0.1$ or less. This prediction should be tested in the future.

\begin{figure}
  \begin{center}
    \FigureFile(80mm,80mm){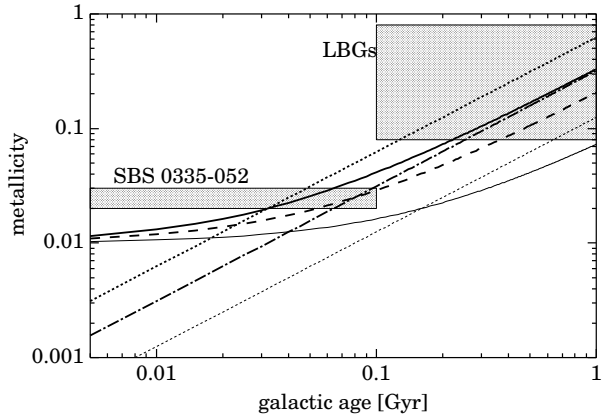}
  \end{center}
  \caption{Metallicity evolution at a galactic age of less than 1
 Gyr. The solid curves are the standard case. The dashed, dotted, and
 dash-dotted curves are the cases of $\eta=0.3$, no infall, and 
 $Z_{\rm in}=0$, respectively. The assumed $\tau_{\rm SF}$ is 1 Gyr and
 5 Gyr for the thick and thin curves, respectively. The shaded areas 
 are the areas indicated from observations of a nearby dwarf galaxy,
 SBS 0335$-$052, and the Lyman break galaxies at $z \sim 3$, as denoted in
 the panel. The vertical axis is normalized by $Z_\odot$. } 
\end{figure}

\begin{figure}
  \begin{center}
    \FigureFile(80mm,80mm){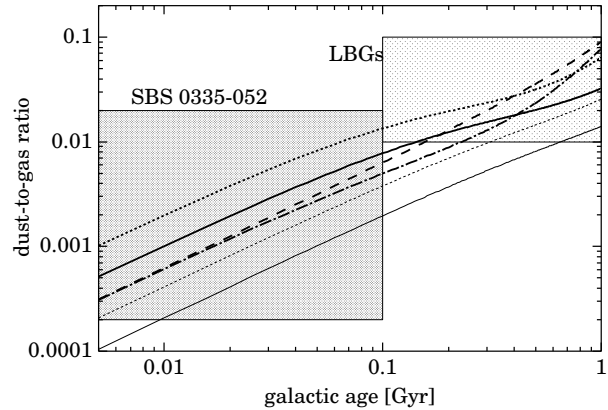}
  \end{center}
  \caption{Dust-to-gas ratio evolution at a galactic age of less than 1
 Gyr. The thick solid, dashed, dash-dotted, and dotted curves are the
 cases of the standard, no SNe destruction, high accretion growth, and
 no infall with  $\tau_{\rm SF}=1$ Gyr, respectively. The thin solid and
 dotted curves are the standard and no infall with $\tau_{\rm SF}=5$
 Gyr, respectively. The effect of metallicity in the infalling gas is
 negligible in this plot. The thick shaded area is the area indicated
 from the observations of SBS 0335$-$052. The thin shaded area is the
 expected area of the Lyman break galaxies at $z \sim 3$. The vertical
 axis is normalized by the dust-to-gas ratio of our Galaxy.}
\end{figure}

For very young galaxies, our model predicts a lower dust-to-metal ratio
than that in \citet{edm01} and \citet{mor03}. This is caused by the
different assumptions between both models, for example, a low or high
accretion efficiency (sticking coefficient), effective or no SNe
destruction, with or without galactic gas infall. However, our smaller
dust-to-metal ratio does not predict more transparent young galaxies
straightforwardly. This is because the amount of attenuation depends
on the dust optical depth, which is strongly affected by the geometry of
the dust distribution. If the dust grains are confined in a small area,
a large attenuation is expected, even in a small dust amount. Of course,
the clumpiness of the distribution is also important.

\section{Conclusion}

We construct a simple evolutionary model of the amount of metal and dust
in order to discuss the evolution of the dust-to-metal ratio
in galaxies. Then, the achieved conclusions are summarized as follows: 

(1) We propose that the grain growth time-scale by the metal
    accretion normalized by the gas metallicity ($\tau_{\rm acc,0}$)
    does not depend on the star-formation history ($\tau_{\rm SF}$)
    significantly. If this is correct, the ratio of the time-scale of dust
    destruction by SNe ($\tau_{\rm SN}$) to that of the accretion growth
    depends on the star-formation history because only $\tau_{\rm SN}$
    depends on $\tau_{\rm SF}$. On the contrary, a constant ratio
    $\tau_{\rm SN}/\tau_{\rm acc,0}$ has been assumed in previous
    studies, such as \citet{dwe98}. 

(2) The evolutionary track in the plane of the metallicity ($Z$) and the
    dust-to-gas ratio ($\cal D$) depend on the ratio $\tau_{\rm
    SN}/\tau_{\rm acc,0}$, or equivalently on the star-formation history,
    $\tau_{\rm SF}$.

(3) The observed linearity of the $Z$--$\cal D$ relation for nearby
    spiral galaxies is interpreted as the sequence of a constant
    galactic age in our framework. 

(4) The dust-to-metal ratio can evolve significantly during the
    evolutionary time-scale of galaxies: the ratio may be 50\%, 30\% and
    20\% of the local value at $z\sim 0.5$, 1, and $\gtsim 2$,
    respectively. Therefore, we will observe $Z$--$\cal D$ relations
    different from the local one for high-$z$ (or young) galaxies.

(5) The uncertainties concerning the time-scales of dust destruction by
    SNe and dust growth by accretion are still large, and
    different parameter choices result in different evolutions of
    the dust-to-metal ratio. Fortunately, we can put a further
    constraint on these time-scales by using the observations of the
    dust-to-metal ratio in high-$z$ (or young) galaxies. Therefore,
    such observations are strongly encouraged.

(6) The effect of infall on evolutionary tracks in the
    $Z$--$\cal D$ plane is negligible. However, since the speed of 
    chemical enrichment depends on the infall rate, the dispersion of
    the infall rate can disturb the $Z$--$\cal D$ relation of a constant
    galactic age.

(7) Our model shows a very good agreement with the observations of a 
    nearby extremely young (age $\ltsim 100$ Myr) dwarf galaxy, SBS
    0335$-$052. We suggests that the Lyman break galaxies at $z \sim 3$
    have a low dust-to-metal ratio of 0.1, or less. Although our model
    predicts a relatively smaller amount of dust in high-$z$ galaxies,
    it does not mean the transparent high-$z$ galaxies, because the dust
    opacity strongly depends on the geometry of the dust distribution.

\bigskip
The author thanks the referee, Dr.\ M.\ G.\ Edmunds, for his great
efforts to understand this work, and also thank Ryuko Hirata and
Hideyuki Kamaya for continuous encouragement. This work is financially
supported by the Research Fellowships of the Japan Society for the
Promotion of Science for Young Scientists.

\end{document}